# Highly Sensitive Ratiometric Fluorescent Fiber Matrixes for Oxygen Sensing with Micrometer-Spatial Resolution

*Giuliana Grasso,[1,§] Valentina Onesto,[1,§] Stefania Forciniti,[1] Eliana D'Amone,[1] Francesco Colella,[1,2] Lara Pierantoni,[3,4] Valeria Famà,[1] Giuseppe Gigli,[1,2] Rui L. Reis,[3,4] Joaquim Miguel Oliveira,[3,4,*] and Loretta L. del Mercato[1,*]*

[1] Institute of Nanotechnology – NANOTEC, Consiglio Nazionale delle Ricerche (CNR), c/o Campus Ecotekne, Via Monteroni, Lecce 73100, Italy;

[2] Department of Mathematics and Physics ''Ennio De Giorgi'', University of Salento, Campus Ecotekne, via Monteroni, 73100, Lecce, Italy;

[3] 3B's Research Group, I3Bs – Research Institute on Biomaterials, Biodegradables and Biomimetics, University of Minho, Headquarters of the European Institute of Excellence on Tissue Engineering and Regenerative Medicine, AvePark, Zona Industrial da Gandra, 4805-017 Barco, Guimarães, Portugal;

[4] ICVS/3B's - PT Government Associate Laboratory, Braga/Guimarães, Portugal.

§: Equally contributing authors
*Corresponding authors: miguel.oliveira@i3bs.uminho.pt; loretta.delmercato@nanotec.cnr.it

**Abstract**

Oxygen-sensing matrixes represent promising tools for live monitoring of extracellular oxygen consumption levels in long term cell cultures. Herein, ratiometric oxygen sensing membranes were prepared by electrospinning, an easy, low cost, scalable and robust method to fabricate nanofibers. Polycaprolactone and poly(dimethylsiloxane) polymers were blended with tris (4,7-diphenyl-1,10-phenanthroline) ruthenium (II) dichloride, used as the oxygen sensing probe, and rhodamine B isothiocyanate, used as the reference dye. The functionalized scaffolds were morphologically characterized by Scanning Electron Microscopy and their physical-chemical profile was obtained by Fourier-Transform Infrared spectroscopy, Thermogravimetric Analysis, and Water Contact Angle. The sensing capabilities were deeply investigated by Confocal Laser Scanning Microscopy, performing photobleaching, reversibility and calibration curves studies towards different concentrations of dissolved oxygen. The electrospun sensing nanofibers showed high response to changes of dissolved oxygen concentrations in the physio-pathological range from 0.5% to 20% and good stability under ratiometric imaging. In addition, the sensing systems were highly biocompatible for cell growth promoting adhesiveness and growth of three cancer cell lines, namely metastatic melanoma cell line SK-MEL2, breast cancer cell line MCF7, and pancreatic ductal adenocarcinoma cell line Panc1, thus recreating *in vitro* suitable biological environment. These oxygen-sensing biomaterials show the potential for measuring alterations in cell metabolism, caused by changes in ambient oxygen content, during either drug testing/validation and tissue regeneration processes.

## 1. Introduction

Hypoxia, a condition in which a tissue goes gradually towards reduced levels of oxygen ($O_2$), is generally recognized as a prodromic event in cancer growth and progression. The master regulator driving to hypoxia is the hypoxia-inducible factor 1 (HIF-1),[1] which is usually silent in normoxic conditions (5-10% $O_2$), whereas it is strongly upregulated during anoxia (less than 1.5% $O_2$) in several cancer types, such as melanoma,[2] breast cancer,[3] and pancreatic ductal adenocarcinoma.[4] The pharmacological and clinical relevance of the $O_2$ concentration make it a well-recognized cancer biomarker.[5]

Over the last years, many strategies have been developed for sensing and imaging $O_2$.[6] Among these, optical oxygen sensors have become attractive thanks to various features, such as reversibility, selectivity and miniaturization. Moreover, optical sensors allow to perform non-invasive measurements and to image $O_2$ at high spatial resolution (micrometer scale).[7, 8] In the last years, many transition metal complexes ($Ru^{2+}$, $Os^{2+}$, $Ir^{3+}$) and metalloporphyrins ($Pt^{2+}$ and $Pd^{2+}$) have been used for producing optical oxygen sensors because of their capability to be dynamically quenched in presence of dissolved oxygen (DO).[9] Combining these oxygen-sensing fluorescent probes to non-sensitive oxygen fluorophores, ratiometric optical systems, that measure fluorescence signals at two different wavelengths, can be produced. Ratiometric systems overcome some sensitivity limitations usually encountered in traditional optical sensors that measure fluorescence intensity only at a single wavelength (*e.g.*, interference from a analyte-independent factors, including instrumental parameters, local microenvironment, probe's concentration, and photobleaching).[10–12] Ratiometric $O_2$-sensing systems based on semiconducting quantum dots, nano- and microparticles, electrospun fibers prepared with natural[13] and synthetic polymers,[14] have been successfully produced for monitoring intracellular concentration changes of $O_2$.[15, 16] For example, Xu *et al.* reported dual-emissive ratiometric nanoscale metal–organic frameworks (NMOFs) for intracellular quantification of oxygen showing that NMOFs are uptake by mouse colon carcinoma CT26 cells allowing reliable measurements of $O_2$ cellular levels by confocal laser scanning microscopy (CLSM).[17] In a different approach, Wen *et al.* reported a challenging protocol to engineer ratiometric afterglow/fluorescence dual-emissive ratiometric $O_2$ polystyrene nanoparticles to explore the hypoxia environment in solid tumors when subcutaneously injected in mice bearing mouse sarcoma cell-derived S180.[18] Also, Gkika *et al.* designed poly-L-lysine-coated polystyrene particles encapsulating a new lipophilic and oxygen-responsive Ru(II) tris-heteroleptic polypyridyl complex together with a reference BODIPY dye to sense normoxic and oxygen deprived (hypoxia) conditions following uptake in A549 lung carcinoma and HeLa cells.[19] Zhao *et al.* incorporated a highly efficient hydrophobic oxygen probe PtTFPP conjugated to a reference fluorescence resonance energy transfer (FRET) donor inside the core of micelles to measure oxygen concentration in HeLa cells.[20] Sensing extracellular changes of $O_2$ is fundamental since it is associated with extracellular matrix remodeling and enhanced invadopodia and metastatic invasion.[21, 22] However, so far, few are the examples reported in literature and regarding the production of diverse $O_2$ sensors based on the ratiometric approach to measure extracellular changes of $O_2$. For instance, Xue *et al.*, demonstrated extracellular detection of $O_2$ changes by producing electrospun core-shell poly(ε-caprolactone)/poly(dimethyl)siloxane (PCL/PDMS) nanofibers embedding $O_2$-quenching ruthenium complexes, or platinum metalloporphyrin dyes, in the PDMS core. The produced fluorescent fiber matrices, based on a single wavelength detection approach, showed oxygen sensing properties and biocompatibility with glioma and glioma-derived primary cells.[23]

In this study, we proposed an original protocol for the fabrication of PCL/PDMS nanofibers embedding the $O_2$-sensing $Ru(dpp)_3^{2+}$ probe and the $O_2$ non-sensitive dye RBITC for ratiometric sensing of microenvironment changes of dissolved oxygen (DO) (**Figure 1**). The homogenous blending of PDMS, PCL, $Ru(dpp)_3^{2+}$ and RBITC, was employed in the electrospinning technique and fibers were obtained by a single ratiometric fluorescent polymeric solution, which resulted in a low-cost and time-saving fabrication method in comparison to the previously reported core-shell set-up.[23, 24] Moreover, the continuous ratiometric fluorescent polymer phase enabled the direct exposure of the oxygen-sensing dye ($Ru(dpp)_3^{2+}$) together with the oxygen-permeable material (PDMS) not only within the lumen of the fibers, but also at the interface with the surrounding environment, thus avoiding their confinement in the inner core of the fibers enclosed by a non-oxygen permeable polymer shell. In addition, we optimized ratiometric systems for sensing oxygen, whose fluorescent signals are not affected by analyte-independent factors[12] usually encountered in fibrous platforms based on single wavelength optical methods. Thanks to their excellent cytocompatibility, these ratiometric oxygen-sensing biomaterial represent a promising platform for future applications in live monitoring of $O_2$ consumption in biological environments and tissue engineering applications.

## 2. Materials and Methods
### 2.1 Materials

Rhodamine B isothiocyanate (RBITC, CAS Number: 36877-69-7) mixed isomers and polycaprolactone (PCL, # 440744) ($M_{av}$=80.000) were purchased from Sigma-Aldrich (Merck KGaA, Darmstadt, Germany). Sylgard®184, two-part silicone elastomer kit (PDMS) was purchased from Dow Corning Corporation (Midland, Michigan, USA). Tris(4,7-diphenyl-1,10-phenantroline)ruthenium(II) dichloride ($Ru(dpp)_3^{2+}$ $Cl_2^-$ (CAS Number: 36309-88-3) was purchased from Alfa Aesar by Thermo Fisher Scientific (Haverhill, Massachusetts, USA). Methanol (Reag. USP, Ph. Eur., for analysis, ACS, ISO) (MeOH, CAS Number: 67-56-1), chloroform (Reag. USP, Ph. Eur., for analysis, ACS, ISO) ($CHCl_3$, CAS Number: 67-66-3), acetone (Reag. USP, Ph. Eur., for analysis, ACS, ISO) (CAS Number: 67-64-1), 2-propanol (Reag. USP, Ph. Eur., for analysis, ACS, ISO) (IPA, CAS Number: 67-63-0), and toluene (Reag. USP, Ph. Eur., for analysis, ACS, ISO) (CAS Number: 108-88-3) were purchased from PanReac AppliChem (Milan, Italy). The N,N-dimethylformamide (anhydrous, 99.8%) (DMF, CAS Number: 68-12-2) was purchased from Sigma-Aldrich (Merck KGaA, Darmstadt, Germany). Tetrahydrofuran (EMSURE® ACS, Reag. Ph. Eur., for analysis) (THF, CAS Number: 109-99-9) was purchased from VWR Chemicals (Milan, Italy).

### 2.2 Fabrication of ratiometric $O_2$-sensing fibers

For electrospinning of ratiometric $O_2$-sensing fibers, PCL/PDMS blends containing dissolved $Ru(dpp)_3^{2+}$ and RBITC fluorophores were mixed in a solvent mixture and gently stirred in dark at room temperature to obtain a homogeneous phase. PCL solutions 12% w/v were prepared 24 hours in advance to allow full dissolution and homogenization of the polymer. To prepare PDMS, elastomer and curing agent were weighted in 10:1 ratio, mixed and degassed in a vacuum pump for 20 minutes. The fluorophores, Ru(dpp)$_3^{2+}$ and RBITC, were prepared in a concentration of 1 mg/mL in each solvent system used in the study. Therefore, the electrospinning solution was composed in the following ratio: 1 part of PDMS w/w, 9 parts of PCL 12% w/v, and a total amount of dyes representing the 0.10% w/v of the polymeric formulation. The solution was left stirring (300 rpm) at room temperature (R.T.) for 2 hours before electrospinning, in the dark. For the electrospinning process, the solution was transferred into a 1 mL syringe (Henke Sass Wolf, Tuttlingen, Germany) equipped with a 21-Gauge stainless steel blunt needle (Sterican®, B-Braun, Milan, Italy). The syringe was placed on a syringe pump (E-fiber, SKE Research Equipment®, Bollate, Italy) and the electrospun fibers were deposited directly onto the surface of squared glasses (10x10 mm) positioned on a stationary collector. The optimized parameters were the following: flow rate 0.8 mL/hour; voltage: 14-6 kV; tip-collector plate distance: 20 cm; time of deposition: 1 minute. Environment conditions were as follow: Temperature range from 25°C to 30°C, and relative humidity from 10% to 20%. The electrospun fiber mat was finally exposed to thermal polymerization process[25] by transferring the mat in a laboratory oven set up at 45°C, for 12 hours. Samples were stored in dark and at room temperature, in a vacuum desiccator until further use.

### 2.3 Characterization of ratiometric O$_2$-sensing fibers

The morphology of ratiometric O$_2$-sensing fibers was analyzed by scanning electron microscopy (SEM, ZEISS Sigma 500, Carl Zeiss, DE). Before imaging, the fibers were sputtered-coated (Compact Coating Unit CCU-010, SafeMatic GmbH, Zizers, Switzerland) with 10 nm of gold (Target Au Ø 54x0.2 mm, purity 99.99%). Images were acquired using an accelerating voltage of 5 kV, secondary electron detector (SE2), with magnifications of 2500x, 5000x, and 10000x. The diameters of the fibers were then extracted by drawing linear regions of interest along the minor axis of the fibers in ImageJ software (version 64-bit Java 1.8.0_172).[26] Images of the most representative electrospun fibers are reported in the **Figure S1**.

Thermogravimetric analysis was carried out through a Simultaneous Thermal Analyzer DSC/TGA STD Q600 (Ta Instruments, Waters™, New Castel, DE), and samples were analyzed in an alumina pan, after tare, heating at 10°C/minute, from 25°C up to 800°C, under N$_2$ atmosphere. Data were plotted and analyzed using TA Universal Analysis 2000 software, Version 4.5A (Ta Instruments, Waters™, New Castel, DE).

The chemical groups exposed on the fibers mat were observed using Fourier-Transform Infrared spectroscopy via a FT/IR-6300 type A spectrophotometer (JASCO, Easton, MD, USA), set up in ATR-FTIR mode, with a resolution of 4 cm$^{-1}$, in the range from 500 cm$^{-1}$ to 4000 cm$^{-1}$. Samples were prepared by electrospun deposition on a piece of silicon wafer (previously washed in IPA and acetone and then dried under a flux of N$_2$) and polymerized in a laboratory oven at 45°C, for 12 hours before analysis.

The hydrophilicity of the fibers was calculated using a CAM 200 (KSV Instruments Ltd., Finland) instrument to measure static water contact angles (WCA) ($\theta$). For the experiment set up, the liquid (heavy phase) was represented by water (d= 0.9986 g/cm$^3$) and the liquid (light phase) was air (d=0.0013 g/cm$^3$). The volume of the drop employed for the measurements and charged on the tip of the needle was of 5 µL. The WCA was calculated using Laplace-Young fitting and the data reported are the average of three measurements performed onto different areas of each sample.

To functionalize the fibers, a Tergeo Plus Plasma Cleaner (PIE Scientific LLC, Union City, CA, USA) instrument was employed. The parameters adopted were as follow: 50% O$_2$ flow rate, 50% power, for 0.3 minutes.

In order to measure the sensing performance of the ratiometric O$_2$-sensing fibers to DO by CLSM imaging, fiber's samples were placed in a 4-well Ibidi® chamber plate and 400 µL of DDW were added in each well. Calibrations were performed via CLSM (SP8 Leica Microsystem, Manheim, Germany) in the oxygen concentration range of 0.5-20%, obtained by purging N$_2$ directly into DDW) while measuring the oxygen content into water by a commercial oxygen meter (Vernier Go Direct® Optical Dissolved Oxygen Probe, Vernier Science Education, Beaverton, Oregon, USA), ~30 minutes time was set to ensure that equilibrium was reached prior to measurement. Images were acquired by using a HC PL FLUOTAR 20x/0.50 dry objective and 2.5 zoom (232.5 µm x 232.5 µm). The emission of Ru(dpp)$_3^{2+}$ and RBITC dyes was sequentially collected at $\lambda_{em}$=550-650 nm ($\lambda_{exc}$=405 nm) and $\lambda_{em}$=570-620 nm ($\lambda_{exc}$=561 nm), respectively. The CLSM images were processed with a custom algorithm written in GNU Octave (version 6.2.0),[27–29] that was opportunely modified in order to automatically quantify the fibers fluorescence intensities for ratiometric analyses. Briefly, the RBITC channel images were first converted to grayscale and binarized. Then, morphological opening with a disk-shaped element was performed to remove any small white noises in the image, and morphological closing to remove any small holes in the object. In this way, fibers were identified in the binarized reference channel image, which was used as a mask to store pixel locations and corresponding fluorescence intensities $I_{Ru(dpp)_3^{2+}}/I_{RBITC}$ (obtained as the pixel-by-pixel ratio between fluorescence intensities of the original $Ru(dpp)_3^{2+}$ and RBITC images, respectively). Mean and standard deviation from 4 different images were finally extracted.

### 2.4 Ratiometric measurements

For the development of our sensing fibrous scaffold, $Ru(dpp)_3^{2+}$ and RBITC were employed as, respectively, an oxygen sensing reporter and a non-sensitive oxygen dye for ratiometric analyses. While RBITC does not respond to oxygen concentration changes,[30] ruthenium complexes oxygen quenching mechanism is extensively reported in literature.[31, 32] The linear dependence of the emission intensity behavior of $Ru(dpp)_3^{2+}$ in correlation with the quencher concentration is governed by the Stern-Volmer equation *(1)*,[33] reported as follows:

$$\frac{I_0}{I} = 1 + K_{sv}[O_2] \quad (1)$$

where, $I_0$ and $I$ are respectively fluorescence intensities in absence and in presence of the quencher; $K_{sv}$ is the Stern–Volmer quenching constant; and $[O_2]$ is the gaseous or dissolved oxygen concentration. In ideal conditions, the Stern-Volmer equation, plotted with the fluorescence intensities against the oxygen concentrations, leads to a linear correlation with a slope equal to $K_{sv}$ and an intercept equal to 1, also providing information on the sensitivity of the sensor in relationship to $O_2$ quenching effect. In our ratiometric oxygen sensing fibers system, the fluorescent response of $Ru(dpp)_3^{2+}$ complex and RBITC towards different concentration of oxygen is derived from the ratio (*R*) between the maximum fluorescence intensity of $Ru(dpp)_3^{2+}$ metal complex and the maximum fluorescence intensity of $O_2$-insensitive RBITC dye, as follows *(2)*:

$$R = \frac{I_{[Ru(dpp)_3^{2+}]}}{I_{RBITC}} \quad (2)$$

The Stern-Volmer equation is then transformed, by the substitution of $I_0/I$ into $R_0/R$, representing the response of the ratiometric sensor, as follows *(3)*:

$$\frac{R_0}{R} = 1 + K_{sv}[O_2] \quad (3)$$

where, $R_0$ is the fluorescent signal in absence of the quencher. In this way, it was possible to evaluate the ratiometric calibration curve of the developed sensing system.

### 2.5 Biocompatibility of ratiometric $O_2$-sensing fibers

The biocompatibility of ratiometric $O_2$-sensing scaffolds was evaluated by *in vitro* cytotoxicity assays using melanoma cell line SK-MEL 2 (HTB-68™, ATCC, Rockville, Md., USA), human pancreatic cancer cells PANC-1 (CRL-1469™, ATCC, Rockville, Md., USA) and breast cancer cells MCF7 (HTB-22™, ATCC, Rockville, Md., USA) cultured at 37 °C in a humidified 5% $CO_2$ incubator. SK-MEL2 cells were routinely cultured in Eagle's Minimum Essential Medium (Sigma-Merck KGaA, Darmstadt, Germany), whereas Panc-1 and MCF7 cells were grown in DMEM medium (Sigma-Merck KGaA, Darmstadt, Germany), both supplemented with 10% Fetal bovine Serum (FBS, Gibco), 2 mM L-glutamine, and 100 U/mL penicillin and streptomycin (Sigma-Merck KGaA, Darmstadt, Germany). Before cell culture experiments, oxygen sensing fibers, deposited on 1x1 $cm^2$ glass slides, were transferred in 24-well plates, sterilized by UV exposure for 30 minutes and functionalized with fibronectin 0.1 mg/mL (Sigma-Merck KGaA, Darmstadt, Germany) for 30 minutes. $4 \times 10^4$ cells/well were seeded in each well and their viability was evaluated through PrestoBlue cell viability Reagent (Thermo Fisher Scientific) at days 1, 3 and 6. Briefly, at each time point, 5 μL of PrestoBlue Reagent was added into each well and incubated for 1 hour at 37°C and 5% $CO_2$. The fluorescent signal was obtained using a microplate reader (ClarioStarPlus, BMG Labtech, DE), at 535 nm of excitation wavelength and 590 nm of emission wavelength. Cells grown directly in the well plate were used as a control.

### 2.6. Statistical analyses

All experiments were performed in triplicate and the results are reported as the mean ± standard error, unless differently stated. Statistical differences were considered significant at $p < 0.05$, using two-way ANOVA analysis. Data analyses and graphing were performed with Microsoft Excel 365 and GraphPad Prism software (v. 8.4.2–2018).

## 3. Results and Discussion
### 3.1 Ratiometric $O_2$-sensing fibers preparation and morphological observations

The fabrication of $O_2$-sensing fibers was successfully assessed by coupling the advantages deriving from two polymers, namely PDMS and PCL. PDMS shows a high oxygen permeability coefficient (P= 695 x $10^{13}$ $cm^3$) resulting from the product of diffusion constant and oxygen solubility coefficient.[34] However, PDMS lacks electrospinnability properties.[35] For this reason, the addition of PCL in the polymeric solution preparation, was evaluated for allowing the electrospinning of monolithic hybrid fibers. Notably, there were two fundamental aspects that influenced the electrospinning of the hybrid polymeric matrix, in particular: the swelling of PDMS in the solvent and the phenomenon of the partition of the solutes between the solvent and PDMS matrix.[36] Focusing on the PDMS swelling, the use of high solubility solvents alone (*e.g.*, chloroform and toluene) is generally not compatible with the fabrication of PDMS-based devices because it causes marked swelling of the polymer. For this reason, it is usually recommended to prepare PDMS using a high solubility solvent mixed with another low solubility solvent (moderately to highly polar, *e.g.*, water, alcohols, amides, sulfoxides) to produce a mixture that does not swell PDMS, or that at least reduces the swelling related to the use of high solubility solvents. As anticipated, the other aspect taken into account during the preparation of the polymer matrix is the partitioning of the fluorophores ($Ru(dpp)_3^{2+}$ and RBITC) in the PDMS phase. This represents a key point for the preparation of the $O_2$-sensitive devices. Therefore, to prepare a homogeneous mixture of the solutes in the composite matrix composed of PCL and PDMS, it is necessary to use a low solubility solvent, capable of solvating both fluorophores, that neither reacts with the fluorophores nor quenches their fluorescence, and that at the same time, does not cause crosslinking of the PDMS-PCL pre-polymer blend.[36] Considering this knowledge, the experimental plan regarding the manufacturing of $O_2$-sensing fibers involved the preparation of a polymeric blend, based on PCL and PDMS, in different couples of solvent systems, (namely DMF:THF, $CHCl_3$:Toluene, MeOH:$CHCl_3$) with the aim of identifying the best combination in terms of uniformity of the solution and good processability (**Table 1**).

**Table 1: Details of polymeric solution preparation to assess electrospinning process and fibers fabrication.** Morphology based on SEM analyses. All figures and size distribution studies of the electrospun samples are reported in the *Electronic Supporting Information* (**Figure S1**).

| Sample codes | Solvent Mixture | Polymers Concentrations (%) and Polymer Ratio (pr) | Polymerization Temperature (°C) | Morphology |
|---|---|---|---|---|
| F1A | DMF:THF (1:1) | PCL 12% | 45 | fibers |
| F1B | DMF:THF (1:1) | PCL 12% | RT | fibers |
| F2A | MeOH:CHCl$_3$ (1:1) | PCL 12% | 45 | fibers |
| F2B | MeOH:CHCl$_3$ (1:1) | PCL 12% | RT | fibers |
| F3A | DMF:THF (1:1) | PCL 12% + PDMS (pr 5:5) | 45 | film forming |
| F3B | DMF:THF (1:1) | PCL 12% + PDMS (pr 5:5) | RT | film forming |
| F4A | MeOH:CHCl$_3$ (1:1) | PCL 12% + PDMS (pr 5:5) | 45 | film forming |
| F4B | MeOH:CHCl$_3$ (1:1) | PCL 12% + PDMS (pr 5:5) | RT | film forming |
| F5A | DMF:THF (1:1) | PCL 12% + PDMS (pr 6:4) | 45 | film forming |
| F5B | DMF:THF (1:1) | PCL 12% + PDMS (pr 6:4) | RT | film forming |
| F6A | MeOH:CHCl$_3$ (1:1) | PCL 12% + PDMS (pr 6:4) | 45 | film forming |
| F6B | MeOH:CHCl$_3$ (1:1) | PCL 12% + PDMS (pr 6:4) | RT | film forming |
| F7A | DMF:THF (1:1) | PCL 12% + PDMS (pr 7:3) | 45 | fibers |
| F7B | DMF:THF (1:1) | PCL 12% + PDMS (pr 7:3) | RT | fibers |
| F8A | MeOH:CHCl$_3$ (1:1) | PCL 12% + PDMS Sylgard (pr 7:3) | 45 | film forming |
| F8B | MeOH:CHCl$_3$ (1:1) | PCL 12% + PDMS Sylgard (pr 7:3) | RT | film forming |
| F9A | DMF:THF (1:1) | PCL 12% + PDMS (PDMS dil 7:3) | 45 | fibers |
| F9B | DMF:THF (1:1) | PCL 12% + PDMS (PDMS dil 7:3) | RT | fibers |
| F10A | MeOH:CHCl$_3$ (1:1) | PCL 12% + PDMS (PDMS dil 7:3) | 45 | fibers |
| F10B | MeOH:CHCl$_3$ (1:1) | PCL 12% + PDMS (PDMS dil 7:3) | RT | film forming |
| F11A | CHCl$_3$:Toluene (1:1) | PCL 12% + PDMS (PDMS dil 7:3) | 45 | electrospray |
| F11B | CHCl$_3$:Toluene (1:1) | PCL 12% + PDMS (PDMS dil 7:3) | RT | electrospray |
| F12A | DMF:THF (1:1) | PCL 12% + PDMS (PDMS dil 8:2) | 45 | film forming |
| F12B | DMF:THF (1:1) | PCL 12% + PDMS (PDMS dil 8:2) | RT | film forming |

| F13A | MeOH:CHCl$_3$ (1:1) | PCL 12% + PDMS (PDMS dil 8:2) | 45 | film forming |
| F13B | MeOH:CHCl$_3$ (1:1) | PCL 12% + PDMS (PDMS dil 8:2) | RT | film forming |
| F14A | CHCl$_3$:Toluene (1:1) | PCL 12% + PDMS (PDMS dil 8:2) | 45 | electrospray |
| F14B | CHCl$_3$:Toluene (1:1) | PCL 12% + PDMS (PDMS dil 8:2) | RT | electrospray |
| F15 | DMF:THF (1:1) | PCL 12% + PDMS (pr 9:1) + dyes | 45 | - |
| F16 | MeOH:CHCl$_3$ (1:1) | PCL 12% + PDMS (pr 9:1) + dyes | 45 | fibers |
| F17 | CHCl$_3$:Toluene (1:1) | PCL 12% + PDMS (pr 9:1) + dyes | 45 | electrospray |

The first step of our study involved the optimization of the electrospinning settings, including the voltage (+14/-6 kV), the flow rate (0.8 mL/hour), tip of the needle static collector distance (20 cm), as well as atmospheric conditions of temperature and relative humidity (T = 25-30°C, RH = 30-40%). Another important optimized parameter was the polymerization temperature of the composite matrix[37] that had to consider the necessity of: (i) not overcoming the low melting point of PCL, (ii) being optimal to crosslink the blend, (iii) starting and lasting in a range of time which allows to preserve the fibers shape. For these reasons, both room temperature and 45°C, and 12 hours as crosslinking time, were chosen as study conditions. PCL, prepared for this work in the concentration of 12% (w/v), was easily electrospun, resulting in defect-free and quite homogeneous fibers, in both DMF:THF and MeOH:CHCl$_3$, even though fibers left drying at RT resulted plane (**Figure S1, F1B-F2B**) in comparison with the fibers dried in the oven at 45°C (**Figure S1, F1A-F2A**). Then, a study on the polymer ratio was performed. The polymer blend in the solution system CHCl$_3$:Toluene was not electrospun since a phase separation between the two polymers PCL:PDMS was obtained.

An equivalent amount of PCL:PDMS, in the polymer ratio 5:5, determined an emulsion solution and the formation of a film foaming fibers (**Figure S1, F3A-F3B-F4A-F4B**). Similarly, the polymer ratio 6:4 (**Table 1, F5A-F5B-F6A-F6B**) could not be electrospun at all, ascribing this phenomenon to the incompatibility of a high concentration of PDMS, major then 30% of the total polymer blend weight percentage, in the electrospinning technique. In fact, the increase of PCL and the decrease of PDMS in the polymer ratio blend (7:3) formed straight fibers in DMF:THF solvent system (**Figure S1, F7A-F7B**), while a wet glaze could be observed around the central core of the fibers deriving from MeOH:CHCl$_3$ blend (**Figure S1, F8A-F8B**). To enhance the electrospinnability and the morphology of the fibers, PDMS was further diluted. In this step, the solution system CHCl$_3$:Toluene was included in the study, taking into consideration the possibility of reducing the force at the interface of the polymer blend deriving from a diluted system. Therefore, the dilution of PDMS in the 7:3 ratio resulted in a heterogeneous population of fibers with different diameters in DMF:THF (**Figure S1, F9A-F9B**) and in unimodal fibers in MeOH:CHCl$_3$ of 5 µm diameter (**Figure S1, F10A-F10B**), with a more rounded shape when polymerization happened at 45°C. In contrast, more diluted PDMS, in the ratio 8:2, in the final polymer blend resulted in wet fibers with film foaming in both solvent system DMF:THF (**Figure S1, F12A-F12B**) and MeOH:CHCl$_3$ (**Figure S1, F13A-F13B**), due to a non-rapid evaporation of the solvents after fiber deposition. In contrast, the polymer blend in CHCl$_3$:Toluene determined an electrospray phenomenon in the Taylor cone, again with no possibility of obtaining electrospun fibers (**Table 1, F11A-F11B** and **Table 1, F14A-F14B**), in both 7:3 and 8:2 PDMS dilutions. The reason has to be ascribed to the swelling effect of the couple solvents CHCl$_3$:Toluene on PDMS.[36] Considering the film-forming fibers derived from an excess of solvent and its non-fast evaporation process at both room temperature and at 45°C, as last chance three polymer blends were prepared using PCL:PDMS (polymer ratio 9:1) with the addiction of the dyes: (i) a polymer blend in MeOH:CHCl$_3$, resulting in a homogeneous phase (**Table 1, F16**); (ii) a polymer blend in CHCl$_3$:Toluene, obtaining a phase separation (after vortexing and removal of the supernatant, electrospray phenomenon did not allow fibers fabrication) (**Table 1, F17**); (iii) a polymer blend in DMF:THF, that gave fluorophore quenching (for this reason it was excluded from our study) (**Table 1, F15**). Considering the aim of validating a simple composite matrix and feasible electrospinning method to prepare ratiometric oxygen sensing fibers, the polymer blend, represented by MeOH:CHCl$_3$ solvent system, was selected.

Once the polymer ratio and the electrospinning parameters were found and optimized, the influence of dye loading on the electrospinnability of the polymeric solution and morphology of resulting fibers were tested, together with their sensing performances. With this aim, comparative studies on the fabricated scaffolds were carried out in order to find the best concentration of each fluorescent component to obtain a final ratiometric system which guaranteed the successful entrapment of the luminescent probes within the fibers. Stock solutions of Ru(dpp)$_3^{2+}$ and RBITC were prepared in the concentration of 1 mg/mL in MeOH:CHCl$_3$ (1:1) and used for the following study. Here, three concentrations of the reference dye RBITC as well as three concentrations of fluorophore Ru(dpp)$_3^{2+}$ were charged in the composite polymer blend (**Table S1**) and the effect of the probe cargo was detected via CLSM imaging. As expected, the dye loading influenced both the production protocol and the morphology of the scaffolds. Indeed, although the optimized fabrication parameters were applied to all solutions, the electrospinnability resulted perturbed with the appearance of electrospray phenomenon and current instabilities when conditions **RS1**, **RS2**, **RS4** and **RS6** (**Table S1**) were adopted. As it is possible to observe in **Figure S2**, keeping the concentration of the sensing ruthenium-based complex constant (22 µg/mL), the increase cargo of the reference dye RBITC (conditions **RS1** and **RS2**, **Table S1**) determined a change in the pattern of the fibrous scaffold, which appeared non homogeneous in size and shape (**Figure S2a-b**). Moreover, the fluorescence signals of RBITC appeared confined at the border of the electrospun strings, which instead presented a $Ru(dpp)_3^{2+}$-rich core (**Figure S2b**). In addition, the fibers appeared wet and oily, likely due to a slowing-down or an interruption of the curing process of the two PDMS components within the polymeric mixture. These results are in agreement with previous report[23] and did not allow to perform any dissolved oxygen sensing test on the fabricated scaffolds. Thus, lowering the amount of the reference probe from 375 µg/mL to 75 µg/mL (condition **RS3**, **Table S1**) resulted in the co-localization of the ratiometric fluorescent signals that uniformly lighted-up the whole fiber mat (**Figure S2c**) upon laser excitation. Having established the RBITC cargo (75 µg/mL), $Ru(dpp)_3^{2+}$ concentrations were also varied in the polymer solution to identify the optimal indicator loading amount for the ratiometric measurements. At this point, both morphology and dissolved oxygen sensing ability were deeply investigated through confocal imaging analysis. An at least twofold increase of $Ru(dpp)_3^{2+}$ to 90 µg/mL or 360 µg/mL (conditions **RS4** and **RS6**, **Table S1**) determined difficulties in the application of the optimized electrospinning parameters, obtaining fibers losing their form and size in both cases (**Figure S3a** and **Figure S3c**), as previously observed. In addition, in the case of the highest $Ru(dpp)_3^{2+}$ concentration (condition **RS6**, **Table S1)** some fluorescent polymeric clusters, found in the blended electrospun fibers, indicated an excessive dye amount resulting in molecular aggregation (**Figure S3c**). Therefore, an intermediate amount of 180 µg/mL of ruthenium-based complex (condition **RS5**, **Table S1**) within the polymer blend yielded to bead-free, fluorescent and microscaled scaffolds (**Figure S3b**). Results from sensor sensitivity experiments at different dissolved oxygen concentrations are showed in **Figure S3d**. Fluorescent emission responses were collected under the same excitation conditions and analysed via image segmentation. For all the sensing systems **RS4**, **RS5** and **RS6** intensity ratios were linearly correlated to the dissolved oxygen concentrations in the surrounding environment with similar slopes and regression coefficients of the curves. These results are in agreement with our expectations since, although varying the amount of the sensing dye within the polymer solutions, the final scaffolds kept the same sensing ability, being they ratiometric platforms.[12, 38, 39] Clearly, the calibration curves reported higher intensity ratios when the amount of the ruthenium-based probe was increased, while keeping the RBITC concentration constant. Although it appeared that the intensity ratio ($I_{Ru(dpp)_3^{2+}}/I_{RBITC}$) was maximized in correspondence of an excessive cargo of the fluorophores ($\geq 0.10\%_{w/w}$ of the polymer blend), the sensing capacity was slightly reduced. Indeed, a diminished slope of the calibration curve of **RS6**, compared to **RS4** and **RS5** scaffolds, was obtained (**Figure S3d**). Moreover, the high standard error during pixel-by-pixel analysis was a consequence of the non-homogenous distribution of the fluorescent probes within the fibers. Therefore, the fabrication condition **RS5**, with dyes representing $0.10\%_{w/w}$ of the polymeric formulation, was selected as the optimal balance to obtain a ratiometric optical sensing fibrous mat.

The microfibers' morphology, investigated through SEM, revealed randomly oriented fibers, with a continuous, smooth surface and a bead-free structure with a string-like shape. Size distribution analyses evidenced uniform fibers diameters, average value of 1.18 µm ± 0.20 (**Figure 1 a-c**). The uniform distribution of both fluorophores within the lumen of the fibers, visualized using CLSM, confirmed the promising ratiometric sensing platform (**Figure 1 d-f**).

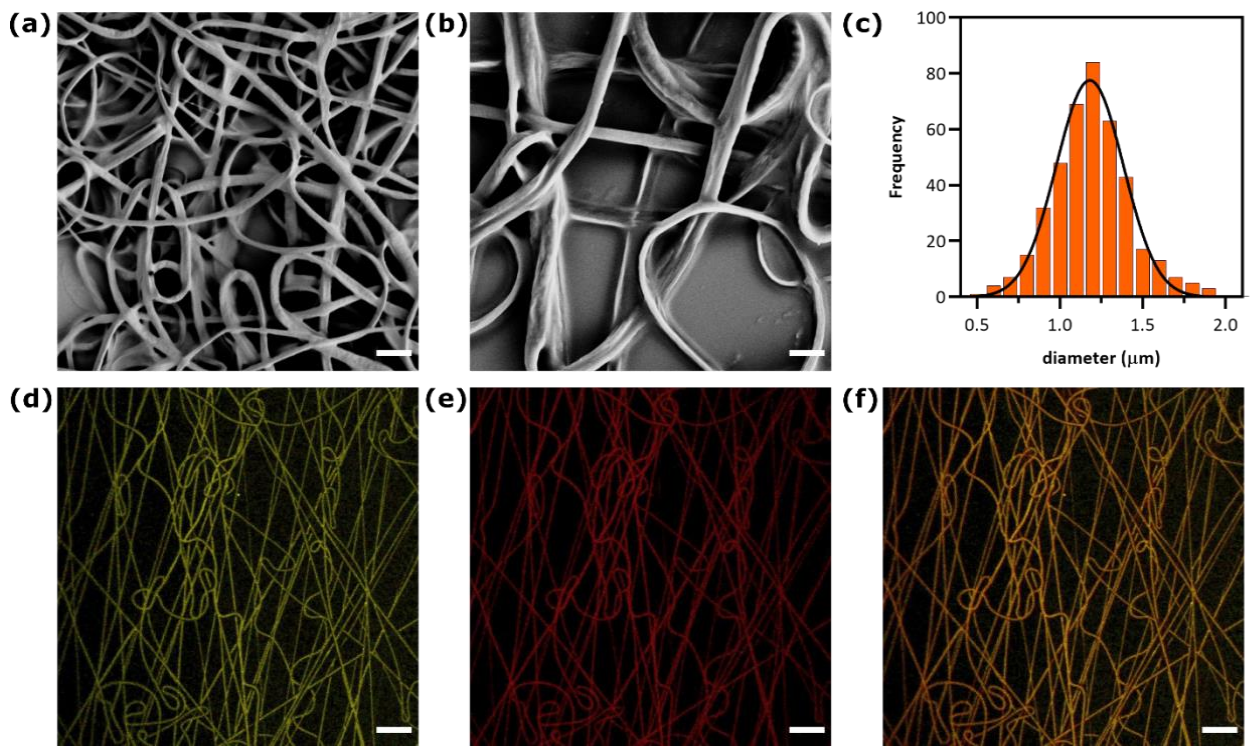

**Figure 1. Morphology of ratiometric $O_2$-sensing fibers. (a)** SEM micrographs of blended PCL/PDMS/Ru(dpp)$_3^{2+}$/RBITC fibers at 2.5x and 5x magnifications; scale bars = 10 µm (left panel); 2 µm (right panel); **(c)** Size distribution studies; and **(d-f)** Representative CLSM micrographs of PCL/PDMS/Ru(dpp)$_3^{2+}$/RBITC fibers. The individual yellow (false color, Ru(dpp)$_3^{2+}$ $\lambda_{exc}$= 405 nm, $\lambda_{em}$=550-650 nm) and red (RBITC $\lambda_{exc}$ = 561 nm; $\lambda_{em}$=570-620 nm) channels are shown, followed by overlay of the two channels. (SP8, Leica, HC PL FLUOTAR 20x/0.50 dry objective, zoom 2.5, scale bars = 20 µm).

### 3.2 Ratiometric $O_2$-sensing fibers structural characterization

The surface chemistry of the composite fibers was studied by the FT-IR, in ATR mode. The spectrum of blended PCL-PDMS fibers was compared with the spectra of PCL fibers and crosslinked PDMS solution, since no fibers could be obtained from PDMS (**Figure 2a**). From the FT-IR spectrum of the blended fibers, it was possible to identify the typical bands of PCL such as, the carbonyl (C=O) overtone at 3446 cm$^{-1}$, the symmetric CH$_2$ stretching at 2938 cm$^{-1}$, the symmetric CH$_2$ stretching at 2860 cm$^{-1}$, the carbonyl (C=O) stretching at 1725 cm$^{-1}$ together with C–O and C–C stretching, the asymmetric C-O-C stretching at 1290 cm$^{-1}$, and O−C−O stretching at 1185 cm$^{-1}$.[40, 41] The proof of the blending solution came from the several characteristic bands of PDMS such as the -CH$_3$ deformation in Si-CH$_3$ at 1260 cm$^{-1}$, the -C-H bending corresponding to the peaks between 1400 and 1420 cm$^{-1}$, the Si-O-Si stretching at 1100-1000 cm$^{-1}$, and the Si-C rocking in the fingerprint region between 825 and 865 cm$^{-1}$.[37, 42] Then, at 700-500 cm$^{-1}$ it was possible to find C-H stretching bands of adjacent aromatic and heterocyclic systems, typical of the Ru(dpp)$_3^{2+}$ and RBITC. All the bands are reported in **Table S2**.

The thermal properties of the polymer matrices were evaluated by TGA and DSC analyses under nitrogen atmosphere (**Table 2**). As it is well established in literature,[43] the semi-crystalline polymer PCL has a T$_m$ of circa 60°C. In contrast, PDMS is a non-crystalline material and thus it has no melting point. However, the crosslinking process, occurring during the curing step of the microfibers at 45°C, determined a slight shift of the composite matrix glass transition to higher temperature (64.52°C), as shown in DSC curve (**Figure S4**). The results reported in TGA curve (**Figure 2b**) demonstrated that within a temperature range of 350–450°C, PCL is rapidly decomposed with complete weight loss (99.36%). On the other hand, only 47.83% weight loss was observed for PDMS up to 800°C, showing that PDMS had a better thermal stability than PCL. Therefore, the preparation of the composite based on these two polymers resulted in fiber matrixes with higher thermal stability. Notably, the DTG curve confirmed the results obtained by TGA. The blend presented three degradation steps: the first decomposition peak was recorded at circa 350-440°C, and this corresponded to PCL degradation; the second peak, at 440-500°C, was related to the fluorophores Ru(dpp)$_3^{2+}$ and RBITC; while the third one, at 530-630°C, was proper of PDMS breakdown of its crosslinked structure.[44]

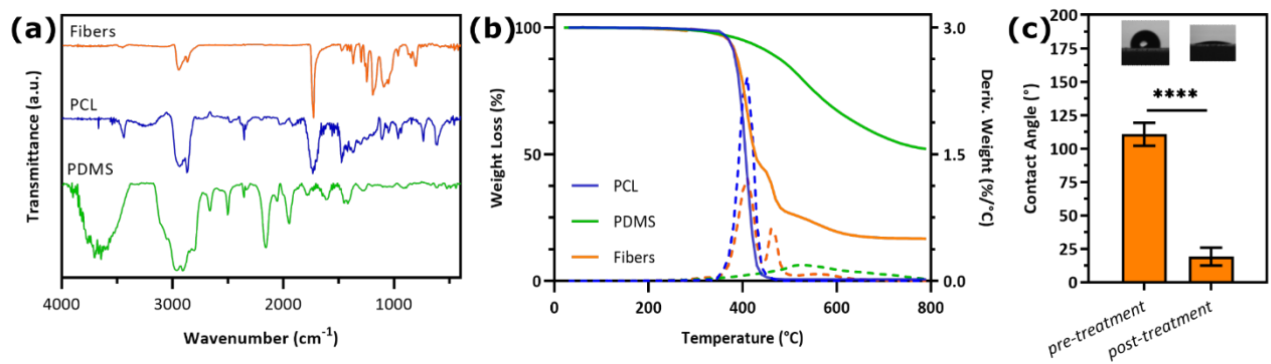

**Figure 2. Structural characterization of ratiometric O$_2$-sensing fibers.** (a) FT-IR spectra of blended PCL/PDMS/Ru(dpp)$_3^{2+}$/RBITC fibers (orange line), PCL (blue line) and PDMS (green line); (b) TGA curves of PCL (straight and dot blue lines), PDMS (straight and dot green lines) and PCL/PDMS/Ru(dpp)$_3^{2+}$/RBITC fibers (straight and dot orange lines); and (c) WCA analysis. The values are reported as mean ± standard deviation (n = 3), p≤0.05(∗), p≤0.0001(∗∗∗∗).

The overall performance of the developed scaffolds stands in the hydrophobicity of the fibrous mat, since this property can obstacle the application of the sensing platform *in vitro* or *in vivo*.[45, 46] Therefore, the optimized ratiometric O$_2$-sensing fibers were subjected to the measurement of the water contact angle, using water as a liquid heavy phase (**Figure 2c**). The drop of water deposited on the polymeric electrospun mats recorded a contact angle θ ≥ 110±7°, indicating a high hydrophobicity. To increase the wettability of the fibers, a functionalization step was carried out with oxygen plasma (50% O$_2$ flow rate, 50% power, for 0.3 minutes), which increased the hydrophilicity of the surface due to the exposure of carboxylic and hydroxyl groups,[45] giving PCL/PDMS/Ru(dpp)$_3^{2+}$/RBITC fibers with a contact angle a θ ≤ 20±6°.

**Table 2: TG analysis of PCL, PDMS and PCL/PDMS/Ru(dpp)$_3^{2+}$/RBITC fibers.**

| Sample | $T_m$ (°C) | $T_i$ decomp. (°C) | $T_{max}$ decomp. (°C) | Weight Loss (%) | Residue (%) |
|---|---|---|---|---|---|
| PCL | 61.59 | 350 | 410 | 99.36 | 0.64 |
| PDMS | / | 460 | 540 | 47.83 | 52.17 |
| PCL/PDMS/Ru(dpp)$_3^{2+}$/RBITC | 64.52 | 360 | 570 | 83.20 | 16.80 |

### 3.3 Ratiometric O$_2$-sensing fibers: dose-response curve studies, photobleaching, and reversibility

One of the main aspects to consider during a real-time dissolved oxygen concentration measurement in living systems is the sensitivity and the stability of the developed analytical platform. As theoretically stated, the luminescence of Ru(II)-based polypyridyl complexes is governed by the metal-ligand charge transfer (MLCT) phenomenon. This latter process determines the transition of an electron from a metal *d* orbital to a ligand $\pi^*$ orbital. As a consequence, upon excitation the exited singlet state is converted to the lowest triplet state through intersystem crossing generating the emissive fluorescence intensities of ruthenium polypyridyl complexes in a sensing system, which in turn is quenched by the presence of oxygen in the close environment.[32] More in detail, two are the main reactions that occur when O$_2$ interacts with Ru(dpp)$_3^{2+}$ complexes thus quenching their fluorescence. The first equation describes the transition of an electron Ru(dpp)$_3^{2+}$ complex (RuL) to give a single oxygen (O$_2^·$) and the ground state of Ru(dpp)$_3^{2+}$ complex, as shown below:

$$RuL_3^{2+·} + O_2 \rightarrow RuL_3^{2+} + O_2^· \qquad (4)$$

The second reaction intervening regards the oxidation of the Ru(dpp)$_3^{2+}$ complex (RuL) operated by the transferring of an electron with subsequent production of the superoxide anion (O$_2^·$-), as indicated following:

$$RuL_3^{2+·} + O_2 \rightarrow RuL_3^{3+} + O_2^- \qquad (5)$$

Therefore, when the 405 nm CLSM laser light excited the Ru(dpp)$_3^{2+}$ probe incorporated in the sensing scaffold, the absorption of light determined the energy gaining of the excited electron which, to restore its ground state, promoted the energy release through the fluorescence emission at higher wavelength. Thus, the fluorescence behaviour of Ru(dpp)$_3^{2+}$ was studied as a function of the dissolved oxygen concentration and resulted in the quenching phenomenon described by the Stern-Volmer equation (see above session 2.4). Even though ruthenium complexes are able to sense dissolved oxygen concentrations alone, as reported in previous published work,[23] however several factors during the analysis (*e.g.:* operational, environmental and instrumental variations) could interfere and change the fluorescence read-out returning inconsistent and non-accurate data. Herein, the fluorescence emission of RBITC, upon excitation with 561 nm CLSM laser light, was not affected by dissolved oxygen concentration. For this reason, the involvement of RBITC as a reference probe coupled to the Ru(dpp)$_3^{2+}$ oxygen sensing ability resulted in a ratiometric analytical method.

Here, the fluorescence response of the ratiometric O$_2$-sensing fibers was measured by recording the fluorescence of the fiber mats (on 1 x 1 cm glass squares) *versus* the DO concentrations in the range 0.5%-20% (**Figure 3a**). **Figure 3a** shows the fluorescence responses of Ru(dpp)$_3^{2+}$ (yellow channel, false color) and RBITC (red channel) deriving from the dissolved oxygen concentration changes in the aqueous medium in which fibers were immersed in order to perform the calibration by sequential CLSM acquisitions. It was evident that the RBITC emission (red channel) was not subjected to any variations in terms of signal intensity, meanwhile the emission of Ru(dpp)$_3^{2+}$ (yellow channel, false color) was strictly correlated to the local O$_2$ concentration, as expected from its photophysical behaviour.[31] Indeed, in an almost hypoxic environment (0.5%), the Ru(dpp)$_3^{2+}$ contained in the fibers emitted a remarkable and strong yellow signal (false color) when excited by 405 nm laser irradiation. Turning towards normoxia values (20%), the Ru(dpp)$_3^{2+}$ signal visibly decreased in intensity, resulting in the clear observation of the lonely RBITC emission (red channel). Then, an automated method was applied for the segmentation and the analysis of CLSM images. Results are reported in **Figure 3b** showing the calibration curve extracted from the relation between fluorescence intensity ratios (I$_{Ru(dpp)_3^{2+}}$/I$_{RBITC}$) and DO levels. The linear fittings obtained plotting both the ratiometric calculation, with a correlation coefficient R$^2$= 0.999 (**Figure 3b**), and the Stern-Volmer equation, with a correlation coefficient R$^2$= 0.963 (**Figure 3c**), confirmed an efficient quenching phenomenon due to the sensing probe Ru(dpp)$_3^{2+}$ in presence of increasing oxygen concentrations in the sample, while RBITC showed a constant fluorescence intensity during the whole experiment, confirming its behavior as a reference fluorophore and therefore not sensitivity to variations in oxygen concentration. To further highlight the performances of the developed scaffold, the ratiometric O$_2$-sensing fibers were tested towards their stability to continuous illumination (photobleaching) by recording the changes of fluorescence emission intensities of Ru(dpp)$_3^{2+}$ and RBITC dyes as function of time (**Figure 3d**). The same area of the sample was illuminated for 10 minutes, recording 1 frame every minute. Photobleaching studies previously published reported that the rate of photobleaching for ruthenium complexes, both in solution and in polymer matrices, could increase by time due to the production of reactive oxygen species (ROS), such as singlet molecular oxygen.[47] However, the PCL/PDMS/Ru(dpp)$_3^{2+}$/RBITC fibers did not undergo a remarkable decay of the fluorescence intensity of the dyes in the tested experimental conditions, thus strengthening the advantages of a ratiometric sensing system. Lastly, the reversibility of the fibers to switches in oxygen concentrations was monitored by applying three consecutive cycles of 0.5%-20% switches. Even in this case, after three switch cycles, the ratiometric calculations was extrapolated to obtain a quantitative data regarding the reversibility test. As shown in **Figure 3e**, the ability of the sensing scaffold in changing rapidly the read-out, from a higher intensity to a lower intensity, confirmed the reversibility of the system and the potentiality of the matrix to monitor in real-time relevant DO levels.

Overall, these results together showed that the hybrid monolithic electrospun fibers could efficiently disperse the Ru(dpp)$_3^{2+}$/RBITC fluorophores, allowing reliable, rapid and stable ratiometric optical sensing and imaging of dissolved oxygen variations.

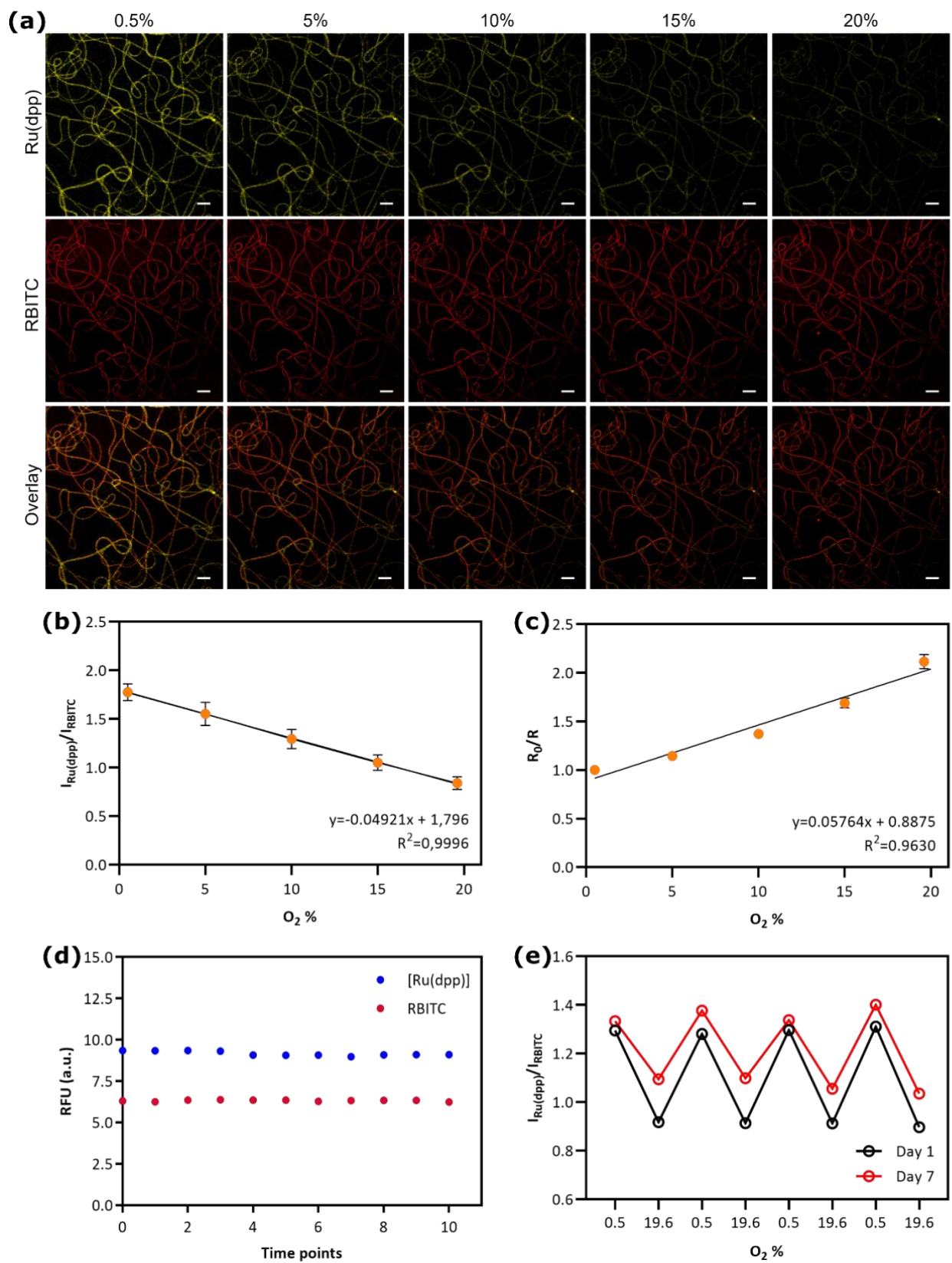

**Figure 3. CLSM imaging and sensing properties of ratiometric $O_2$-sensing fibers. (a)** Representative CLSM micrographs showing the fluorescence response of PCL/PDMS/Ru(dpp)$_3^{2+}$/RBITC fibers *versus* different DO levels (0.5-20%). The individual yellow (false color, Ru(dpp)$_3^{2+}$ $\lambda_{exc}$= 405 nm, $\lambda_{em}$=550-650 nm) and red (RBITC $\lambda_{exc}$= 561 nm; $\lambda_{em}$=570-620 nm) channels of CLSM images are shown, followed by overlay of the two channels. Scale bars = 10 μm. **(b)** $O_2$-calibration curve of the sensing scaffolds as function of oxygen: the fluorescence intensity ratios between the $O_2$-sensitive dye Ru(dpp)$_3^{2+}$ at 500–650 nm ($\lambda_{exc}$: 405 nm) and the reference fluorophore RBITC at 570–620 nm ($\lambda_{exc}$: 561 nm) were plotted *versus* the $O_2$ concentrations obtained fluxing $N_2$ in DDW. Points reported in the plot are the mean value

of three independent readings. Error bars show the standard deviation. **(c)** Stern-Volmer plot of the fibrous ratiometric optical sensing system: the fluorescence ratiometric behaviour of $Ru(dpp)_3^{2+}$ ($\lambda_{exc}$: 405 nm; $\lambda_{em}$: 500-650 nm) and RBITC ($\lambda_{exc}$: 561; $\lambda_{em}$: 570-620 nm) was plotted *versus* the $O_2$ concentrations, obtained fluxing $N_2$ in DDW, to achieve Stern-Volmer curve. **(d)** Photobleaching test: the fluorescence intensity of $Ru(dpp)_3^{2+}$ and RBITC signals were recorded by CLSM, in time lapse mode, illuminating the sample for 10 minutes (1 frame/minute). **(e)** Reversibility of $O_2$-sensing fibers evaluated by conducting sequences of 4 $O_2$ switches between $O_2$ 0% and $O_2$ 19%. Ageing of $O_2$-sensing PCL/PDMS fibers was evaluated by repeating sequences of 4 $O_2$ switches after 7 days on the samples adopted to evaluate the reversibility.

### 3.4 Biocompatibility of ratiometric $O_2$-sensing fibers

The biocompatibility of ratiometric $O_2$-sensing fibers was investigated using three cancer cell lines. To this aim, SK-MEL2, Panc-1, or MCF-7 cells were cultured on a 2D adherent surface (control) and on ratiometric $O_2$-sensing fibers and their viability was measured for up to six days by the PrestoBlue® assay. As shown by the CLSM images reported in **Figure 5**, all tested tumor cells adhered to PCL/PDMS/$Ru(dpp)_3^{2+}$/RBITC fibers retaining their typical epithelial morphology. Notably, cells on the fibrous matrixes formed small groups that tended to cluster together (arrows in **Figures 5a, b** and **c**). Furthermore, cell proliferation was monitored after 1, 3, and 6 days of culture. As reported in **Figure 5d**, no cytotoxicity was observed regarding cell growth on the PCL/PDMS/$Ru(dpp)_3^{2+}$/RBITC fibers as compared to 2D adherent surface. Interestingly, all tested cell lines cultured onto the fibrous scaffolds showed a better growth as compared to that observed for cultures on 2D adherent surfaces (**Figures 5 d, e, f**). These results suggest that the obtained system can recreate a suitable biological microenvironment for tumor cell growth over time.

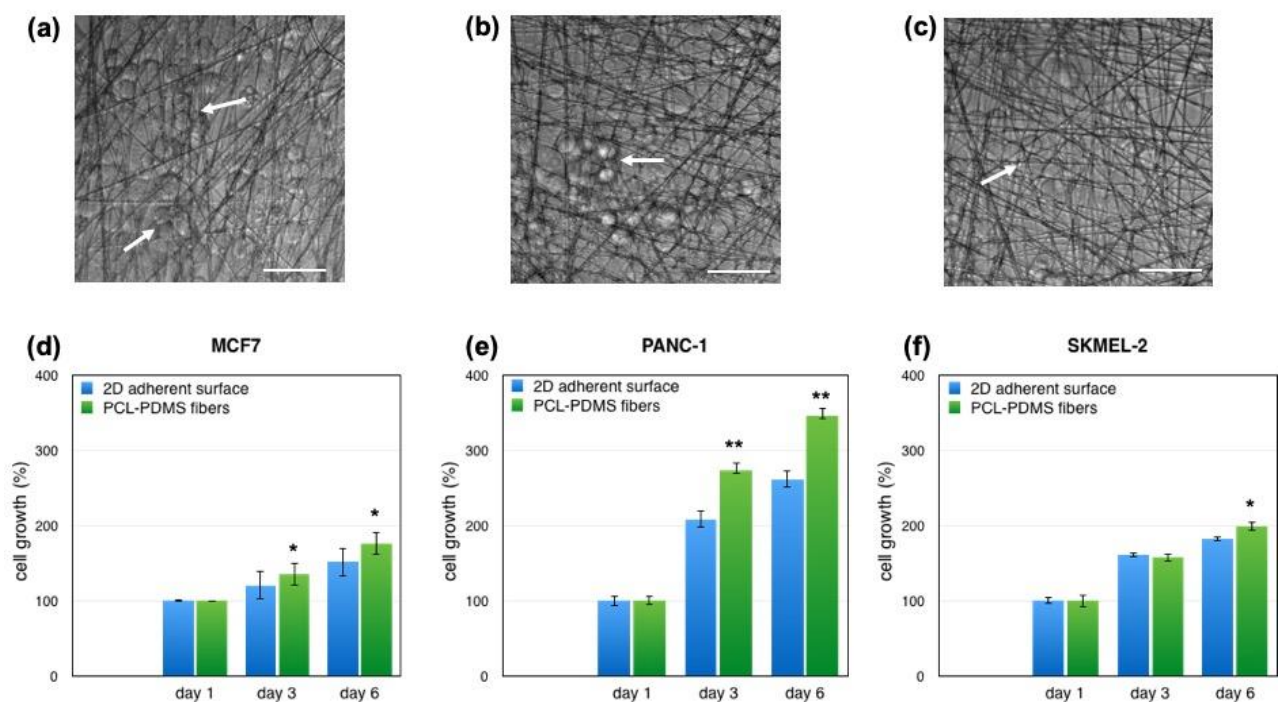

**Figure 4. Biocompatibility and cell proliferation on ratiometric $O_2$-sensing fibers** of **(a, d)** breast cancer cells (MCF-7), **(b, e)** human pancreatic cancer cells (PANC-1), and **(c, f)** malignant melanoma (SK-MEL2) cell lines. Cell viability was measured by PrestoBlue® Reagent as described in Materials and Methods. Scale bars = 50 µm. Values are the means (±SE) of three independent experiments. Statistical analysis: $p<0.05$ (*), $p<0.01$ (**) 2D adherent surface *versus* PCL-PDMS fibers for each time point.

### 4. Conclusions

The oxygen indicator probe $Ru(dpp)_3^{2+}$ and the reference dye RBITC were co-embedded within the lumen of monolithic PCL/PDMS fibers yielding to a hybrid platform for the imaging and ratiometric optical sensing of DO in liquid media. The system combines the properties of the carrier polymer PCL to the transparency and permeability to oxygen of PDMS as well as the simplicity of the set-up configuration for spinning monolithic fibers. Chemical-physical and morphological properties together with stability tests of PCL/PDMS/$Ru(dpp)_3^{2+}$/RBITC fibers were analyzed in detail, revealing the production of bead-free randomly oriented and ratiometric oxygen sensing fibers characterized by a fast, highly reproducible, and stable response to DO over time. Oxygen gradients generated emulating normoxia and hypoxia conditions were detected through 3D (x,y,z) confocal imaging of ratiometric PCL/PDMS/Ru(dpp)$_3^{2+}$/RBITC fibers. Results showed that the emission ratio ($I_{Ru(dpp)_3^{2+}}/I_{RBITC}$) was strictly correlated to the local DO concentration, pointing out a linear sensitivity in the 0.5%-20% range. Interestingly, when the hybrid fibers' response was evaluated during drastic switches in oxygen concentrations, they expressed the ability to rapidly change their read-out, attesting their reversibility and therefore their potential in the detection of DO levels in real-time. These data support the use of such sensing platform for precisely and spatio-temporally monitoring of oxygen extracellular changes and metabolism of cells in *in vitro* or *in vivo* disease models, evaluating how oxygen changes promote or reduce the efficacy of targeted therapies for the treatment of the tumors. Additional sensors (*e.g.*, pH or lactate probes) could be integrated into the lumen of the fibers and their application for non-invasive and accurate *in situ* microenvironment investigation of target analytes could be exploited thanks to the high cytocompatibility of the extracellular matrix-like fiber mats.

**Conflicts of interest**
There are no conflicts to declare.

**Acknowledgements**
The authors acknowledge funding from the European Research Council (ERC) under the European Union's Horizon 2020 research and innovation program ERC Starting Grant "INTERCELLMED" (contract number 759959), the European Union's Horizon 2020 research and innovation programme under grant agreement No. 953121 (FLAMIN-GO), the Associazione Italiana per la Ricerca contro il Cancro (AIRC) (MFAG-2019, contract number 22902), the "Tecnopolo per la medicina di precisione" (TecnoMed Puglia) - Regione Puglia: DGR n.2117 of 21/11/2018, CUP: B84I18000540002), the Italian Ministry of Research (MUR) under the complementary actions to the NRRP "Fit4MedRob" Grant (PNC0000007, CUP B53C22006960001) and "ANTHEM" Grant (PNC0000003); the "NFFA-DI" Grant (B53C22004310006) co-funded by NextGenerationEU and the PRIN 2022 (2022CRFNCP, CUP B53C22006710001) funded by European Union – Next Generation EU. The authors also thank the financial support provided under the project "HEALTH-UNORTE: Setting-up biobanks and regenerative medicine strategies to boost research in cardiovascular, musculoskeletal, neurological, oncological, immunological and infectious diseases", reference NORTE-01-0145-FEDER-000039, funded by the Norte Portugal Regional Coordination and Development Commission (CCDR-N), under the NORTE2020 Program.
The authors gratefully thank Dr. Anil Chandra for initial support (Centre for Research in Pure and Applied Sciences, Bangalore, India) and Dr. Alessandra Quarta (Institute of Nanotechnology, Lecce, Italy) for providing breast cancer cells MCF7. S.F. also acknowledge the AIRC Short-term Fellowship programme.

**References**

1. Kim LC, Simon MC (2022) Hypoxia-Inducible Factors in Cancer. Cancer Res 82:195–196. https://doi.org/10.1158/0008-5472.CAN-21-3780

2. D'Aguanno S, Mallone F, Marenco M, Del Bufalo D, Moramarco A (2021) Hypoxia-dependent drivers of melanoma progression. J Exp Clin Cancer Res 40:159. https://doi.org/10.1186/s13046-021-01926-6

3. Tang K, Zhu L, Chen J, Wang D, Zeng L, Chen C, Tang L, Zhou L, Wei K, Zhou Y, Lv J, Liu Y, Zhang H, Ma J, Huang B (2021) Hypoxia Promotes Breast Cancer Cell Growth by Activating a Glycogen Metabolic Program. Cancer Res 81:4949–4963. https://doi.org/10.1158/0008-5472.CAN-21-0753

4. Tao J, Yang G, Zhou W, Qiu J, Chen G, Luo W, Zhao F, You L, Zheng L, Zhang T, Zhao Y (2021) Targeting hypoxic tumor microenvironment in pancreatic cancer. J Hematol Oncol 14:14. https://doi.org/10.1186/s13045-020-01030-w

5. Sebestyén A, Kopper L, Dankó T, Tímár J (2021) Hypoxia Signaling in Cancer: From Basics to Clinical Practice. Pathol Oncol Res 27:1–15. https://doi.org/10.3389/pore.2021.1609802

6. Roussakis E, Li Z, Nichols AJ, Evans CL (2015) Oxygen-Sensing Methods in Biomedicine from the Macroscale to the Microscale. Angew Chemie Int Ed 54:8340–8362. https://doi.org/10.1002/anie.201410646

7. Yoshihara T, Hirakawa Y, Hosaka M, Nangaku M, Tobita S (2017) Oxygen imaging of living cells and tissues using luminescent molecular probes. J Photochem Photobiol C Photochem Rev 30:71–95. https://doi.org/10.1016/j.jphotochemrev.2017.01.001

8. Wang X, Wolfbeis OS (2014) Optical methods for sensing and imaging oxygen: materials, spectroscopies and applications. Chem Soc Rev 43:3666–3761. https://doi.org/10.1039/C4CS00039K

9. Grist SM, Bennewith KL, Cheung KC (2022) Oxygen Measurement in Microdevices. Annu Rev Anal Chem 15:221–246. https://doi.org/10.1146/annurev-anchem-061020-111458

10. Raju L, Rajkumar E (2023) Coordination compounds of iron, ruthenium and osmium. In: Photochemistry and Photophysics of Coordination Compounds. Elsevier, pp 135–203

11. Gui R, Jin H, Bu X, Fu Y, Wang Z, Liu Q (2019) Recent advances in dual-emission ratiometric fluorescence probes for chemo/biosensing and bioimaging of biomarkers. Coord Chem Rev 383:82–103. https://doi.org/10.1016/j.ccr.2019.01.004

12. Bigdeli A, Ghasemi F, Abbasi-Moayed S, Shahrajabian M, Fahimi-Kashani N, Jafarinejad S, Farahmand Nejad MA, Hormozi-Nezhad MR (2019) Ratiometric fluorescent nanoprobes for visual detection: Design principles and recent advances - A review. Anal Chim Acta 1079:30–58. https://doi.org/10.1016/j.aca.2019.06.035

13. Feng Q, Zhang K, Yang B, Yu Y (2022) Editorial: Biomedical applications of natural polymers. Front Bioeng Biotechnol 10:1–3. https://doi.org/10.3389/fbioe.2022.1077823

14. Wei Q, Deng N-N, Guo J, Deng J (2018) Synthetic Polymers for Biomedical Applications. Int J Biomater 2018:1–2. https://doi.org/10.1155/2018/7158621

15. Chandra A, Prasad S, Gigli G, del Mercato LL (2020) Fluorescent nanoparticles for sensing. In: Colloids for Nanobiotechnology, 1st ed. Elsevier Ltd., pp 117–149

16. Grasso G, Colella F, Forciniti S, Onesto V, Iuele H, Siciliano AC, Carnevali F, Chandra A, Gigli G, del Mercato LL (2023) Fluorescent nano- and microparticles for sensing cellular microenvironment: past, present and future applications. Nanoscale Adv 5:4311–4336. https://doi.org/10.1039/D3NA00218G

17. Xu R, Wang Y, Duan X, Lu K, Micheroni D, Hu A, Lin W (2016) Nanoscale Metal–Organic Frameworks for Ratiometric Oxygen Sensing in Live Cells. J Am Chem Soc 138:2158–2161. https://doi.org/10.1021/jacs.5b13458

18. Wen Y, Zhang S, Yuan W, Feng W, Li F (2023) Afterglow/Fluorescence Dual-Emissive Ratiometric Oxygen Probe for Tumor Hypoxia Imaging. Anal Chem 95:2478–2486. https://doi.org/10.1021/acs.analchem.2c04764

19. Gkika KS, Kargaard A, Burke CS, Dolan C (2021) Ru(II)/BODIPY core co-encapsulated ratiometric nanotools for intracellular O2 sensing in live cancer cells. RSC Chem Biol 18:1520–1533. https://doi.org/https://doi.org/10.1039/D1CB00102G

20. Zhao Q, Pan T, Xiang G, Mei Z, Jiang J, Li G, Zou X, Chen M, Sun D, Jiang S, Tian Y (2018) Highly efficient ratiometric extracellular oxygen sensors through physical incorporation of a conjugated polymer and PtTFPP in graft copolymers. Sensors Actuators B Chem 273:242–252. https://doi.org/10.1016/j.snb.2018.06.026

21. Murphy DA, Courtneidge SA (2011) The "ins" and "outs" of podosomes and invadopodia: characteristics, formation and function. Nat Rev Mol Cell Biol 12:413–426. https://doi.org/10.1038/nrm3141

22. Shin DS, Schroeder ME, Anseth KS (2022) Impact of Collagen Triple Helix Structure on Melanoma Cell Invadopodia Formation and Matrix Degradation upon BRAF Inhibitor Treatment. Adv Healthc Mater 11:. https://doi.org/10.1002/adhm.202101592

23. Xue R, Behera P, Xu J, Viapiano MS, Lannutti JJ (2014) Polydimethylsiloxane core–polycaprolactone shell nanofibers as biocompatible, real-time oxygen sensors. Sensors Actuators B Chem 192:697–707. https://doi.org/10.1016/j.snb.2013.10.084

24. Xue R, Ge C, Richardson K, Palmer A, Viapiano M, Lannutti JJ (2015) Microscale Sensing of Oxygen via Encapsulated Porphyrin Nanofibers: Effect of Indicator and Polymer "Core" Permeability. ACS Appl Mater Interfaces 7:8606–8614. https://doi.org/10.1021/acsami.5b00403

25. Bardelli T, Marano C, Briatico Vangosa F (2021) Polydimethylsiloxane crosslinking kinetics: A systematic study on Sylgard184 comparing rheological and thermal approaches. J Appl Polym Sci 138:1–12. https://doi.org/10.1002/app.51013

26. Schneider CA, Rasband WS, Eliceiri KW (2012) NIH Image to ImageJ: 25 years of image analysis. Nat Methods 9:671–675. https://doi.org/10.1038/nmeth.2089

27. Rizzo R, Onesto V, Forciniti S, Chandra A, Prasad S, Iuele H, Colella F, Gigli G, del Mercato LL (2022) A pH-sensor scaffold for mapping spatiotemporal gradients in three-dimensional in vitro tumour models. Biosens Bioelectron 212:114401. https://doi.org/10.1016/j.bios.2022.114401

28. Onesto V, Forciniti S, Alemanno F, Narayanankutty K, Chandra A, Prasad S, Azzariti A, Gigli G, Barra A, Martino A De, Martino D De, del Mercato LL (2023) Probing Single-Cell Fermentation Fluxes and Exchange Networks via pH-Sensing Hybrid Nanofibers. ACS Nano 17:3313–3323. https://doi.org/10.1021/acsnano.2c06114

29. Rizzo R, Onesto V, Morello G, Iuele H, Scalera F, Forciniti S, Gigli G, Polini A, Gervaso F, del Mercato LL (2023) pH-sensing hybrid hydrogels for non-invasive metabolism monitoring in tumor spheroids. Mater Today Bio 20:100655. https://doi.org/10.1016/j.mtbio.2023.100655

30. Zeng S, Liu X, Kafuti YS, Kim H, Wang J, Peng X, Li H, Yoon J (2023) Fluorescent dyes based on rhodamine derivatives for bioimaging and therapeutics: recent progress, challenges, and prospects. Chem Soc Rev 52:5607–5651. https://doi.org/10.1039/d2cs00799a

31. J. N. Demas DD and EWH (1973) Oxygen Quenching of Charge-Transfer Excited States of Ruthenum(II) Complexes. Evidence for Singlet Oxygen Production. J Am Chem Soc 95:6864–65

32. Quaranta M, Borisov SM, Klimant I (2012) Indicators for optical oxygen sensors. Bioanal Rev 4:115–157. https://doi.org/10.1007/s12566-012-0032-y

33. Gehlen MH (2020) The centenary of the Stern-Volmer equation of fluorescence quenching: From the single line plot to the SV quenching map. J Photochem Photobiol C Photochem Rev 42:100338. https://doi.org/10.1016/j.jphotochemrev.2019.100338

34. Amao Y (2003) Probes and Polymers for Optical Sensing of Oxygen. Microchim Acta 143:1–12. https://doi.org/10.1007/s00604-003-0037-x

35. Yang D, Liu X, Jin Y, Zhu Y, Zeng D, Jiang X, Ma H (2009) Electrospinning of Poly(dimethylsiloxane)/Poly(methyl methacrylate) Nanofibrous Membrane: Fabrication and Application in Protein Microarrays. Biomacromolecules 10:3335–3340. https://doi.org/10.1021/bm900955p

36. Lee JN, Park C, Whitesides GM (2003) Solvent Compatibility of Poly(dimethylsiloxane)-Based Microfluidic Devices. Anal Chem 75:6544–6554. https://doi.org/10.1021/ac0346712

37. Berean K, Ou JZ, Nour M, Latham K, McSweeney C, Paull D, Halim A, Kentish S, Doherty CM, Hill AJ, Kalantar-zadeh K (2014) The effect of crosslinking temperature on the permeability of PDMS membranes: Evidence of extraordinary $CO_2$ and $CH_4$ gas permeation. Sep Purif Technol 122:96–104.

https://doi.org/10.1016/j.seppur.2013.11.006

38. Huang X, Song J, Yung BC, Huang X, Xiong Y, Chen X (2018) Ratiometric optical nanoprobes enable accurate molecular detection and imaging. Chem Soc Rev 47:2873–2920. https://doi.org/10.1039/C7CS00612H

39. Pei X, Pan Y, Zhang L, Lv Y (2021) Recent advances in ratiometric luminescence sensors. Appl Spectrosc Rev 56:324–345. https://doi.org/10.1080/05704928.2020.1793770

40. Riaz U, Ashraf SM (2014) Characterization of Polymer Blends with FTIR Spectroscopy. In: Characterization of Polymer Blends. Wiley, pp 625–678

41. Fiore T, Pellerito C (2021) Infrared Absorption Spectroscopy. In: Spectroscopy for Materials Characterization. Wiley, pp 129–167

42. Gupta NS, Lee K-S, Labouriau A (2021) Tuning Thermal and Mechanical Properties of Polydimethylsiloxane with Carbon Fibers. Polymers (Basel) 13:1141. https://doi.org/10.3390/polym13071141

43. Guarino V, Gentile G, Sorrentino L, Ambrosio L (2017) Polycaprolactone: Synthesis, Properties, and Applications. In: Encyclopedia of Polymer Science and Technology. John Wiley & Sons, Inc., Hoboken, NJ, USA, pp 1–36

44. Wang Y, Cai Y, Zhang H, Zhou J, Zhou S, Chen Y, Liang M, Zou H (2021) Mechanical and thermal degradation behavior of high-performance PDMS elastomer based on epoxy/silicone hybrid network. Polymer (Guildf) 236:124299. https://doi.org/10.1016/j.polymer.2021.124299

45. Niemczyk-Soczynska B, Gradys A, Sajkiewicz P (2020) Hydrophilic Surface Functionalization of Electrospun Nanofibrous Scaffolds in Tissue Engineering. Polymers (Basel) 12:2636. https://doi.org/10.3390/polym12112636

46. Zupančič Š (2019) Core-shell nanofibers as drug delivery systems. Acta Pharm 69:131–153. https://doi.org/10.2478/acph-2019-0014

47. Fuller ZJ, Bare WD, Kneas KA, Xu W-Y, Demas JN, DeGraff BA (2003) Photostability of Luminescent Ruthenium(II) Complexes in Polymers and in Solution. Anal Chem 75:2670–2677. https://doi.org/10.1021/ac0261707